\documentclass[pre,showpacs,12pt,preprintnumbers,amssymb,amsmath]{revtex4}
\usepackage{graphicx}
\usepackage{color}

\begin{document}

\bibliographystyle{apsrev}

\title{Non-additive properties of finite 1D Ising chains with
long-range interactions}

\author{S. S. Apostolov, Z. A. Mayzelis, O. V. Usatenko,}
\thanks{usatenko@ire.kharkov.ua}
\author{V. A. Yampol'skii} \affiliation{A. Ya. Usikov Institute for
Radiophysics and Electronics \\Ukrainian Academy of Sciences, 12
Proskura Street, 61085 Kharkov, Ukraine}

\begin{abstract}
We study the statistical properties of Ising spin chains with finite
(although arbitrary large) range of interaction between the
elements. We examine mesoscopic subsystems (fragments of an Ising
chain) with the lengths comparable with the interaction range. The
equivalence of the Ising chains and the multi-step Markov sequences
is used for calculating different non-additive statistical
quantities of a chain and its fragments. In particular, we study the
variance of fluctuating magnetization of fragments, magnetization of
the chain in the external magnetic field, etc. Asymptotical
expressions for the non-additive energy and entropy of the
mesoscopic fragments are derived in the limiting cases of weak and
strong interactions.
\end{abstract}
\pacs{05.40.-a, 02.50.Ga, 05.50.+q}

\maketitle

\section{Introduction}

One of the basic postulates of the traditional statistical physics
is the assumption that the range of particle interaction is small
compared to the system size. If this condition does not hold the
internal and free energy, entropy, and other physical quantities are
no longer additive. Due to this fact, there are no straightforward
definitions of the temperature, entropy, etc. The statistics of the
systems with long-range interaction are not Gibbsian; Boltzman's
relationship between the entropy and the statistical weight is no
longer valid; the fluctuations of sums of random quantities do not
obey the Gaussian statistics.

Systems with long-range interaction are common in
physics~\cite{daux}. Amongst them are: gravitating
systems~\cite{padma}, electrically charged systems~\cite{nichol},
magnets with dipolar interactions~\cite{barre}, etc. An important
physical example of such class of systems is the \emph{Ising spin
chain} with long-range interaction (see, e.g.,~\cite{isi}).

The long-range interaction leads to the correlations between the
elements of the system, which is the reason for the appearance of
non-Gaussian statistics. Such non-Gaussian distributions (induced by
the long-range correlations) are found in various fields of science,
e.g., in linguistics (distribution of ``words'' in a given
text~\cite{zipf,muyag,KUYa-05}), biology (the distribution of
nucleotides in DNA sequences~\cite{DNA}), computer science (forward
error correction codes~\cite{code}), economics (the distributions in
financial markets~\cite{mant}), sociology, physiology, seismology.

One of the most interesting phenomena in the long-range correlated
systems is the phenomenon of phase transition. The existence of
phase transitions is of great importance for different sciences, for
example, biology in the problem of DNA melting~\cite{pol,dna1}.
Dyson~\cite{dyson}, Ruelle~\cite{rue}, Thouless~\cite{thou} found
different conditions for existence of the phase transition in the
Ising chain with the infinite scale of interaction.

The usual objects for studying the effects of long-range
correlations are the non-extensive thermodynamic
systems~\cite{tsallis} where both interaction range and the system
length are macroscopically large. However, the \emph{mesoscopic}
long-range correlated systems, which are not large enough to
consider the thermodynamic limit, have not been studied thoroughly.
The statistical properties of such systems are very similar to the
properties of the infinite \emph{non-extensive thermodynamic
systems}. In the present paper, we consider the \emph{finite-size
subsystems} (fragments) of an \emph{infinite Ising chain} as
examples of mesoscopic correlated systems. We study the Ising chain
with the Hamiltonian,
\begin{equation}\label{hamilt}
\mathcal{H}=-\sum\limits_{0<j-i<N}\varepsilon(j-i)s_i
s_j-H\sum\limits_{i}s_i ,
\end{equation}
where the spin variable $s_i$ takes on two values, $-1$ and $1$,
$\varepsilon(r)>0$ is the exchange integral of the ferromagnetic
coupling, and $H$ is the external magnetic field. Below, for the
simplicity, we assume $\varepsilon(r)=0$ for $r>N$. The range $N$ of
spin interaction is finite although arbitrary. The chain is assumed
to be in the thermodynamical equilibrium with the Gibbs thermostat
of temperature $T$ (unless otherwise mentioned). Such a system is
obviously extensive one, and the usual thermodynamics can be applied
to the system \emph{as a whole}. There is no spontaneous
magnetization in such a chain due to the finiteness of the particle
interaction range. Nevertheless, its \emph{mesoscopic subsystems}
are of interest because of their non-additive statistical
properties. The aim of our work is to calculate different
statistical quantities for an mesoscopic subsystem of length $L$,
that can be of the order of the interaction length $N$.

Thus, we study an Ising chain with the finite range of interaction,
but the lengths of the subsystems \emph{are also finite}. We show
that energy and entropy of the fragments of the Ising chain are
non-additive quantities. We calculate the root-mean-square (RMS)
value of fluctuation of the magnetization which does not scale as
$\sqrt{L}$ (the number of spins in a fragment), unlike the case of
short-range interactions. We suggest a way to calculate statistical
quantities for the finite subsystems in contact with both the
external Gibbs thermal bath of temperature $T$ and the rest of the
Ising chain playing a role of additional thermostat.

For mesoscopic systems, an important problem is to introduce the
appropriate statistical ensemble correctly, because different
statistical ensembles are not equivalent for such systems (see,
e.g., Ref.~\onlinecite{caset}). These canonical ensembles are
equivalent in the thermodynamical limit only. In the present paper,
we deal with an ensemble of the finite subsystems that are the parts
of an infinite Ising chain.

We calculate the statistical properties of the Ising chains using
their equivalence with the binary $N$\emph{-step Markov chains}. The
multi-step Markov chain is the sequence where symbols, say $\pm 1$,
are generated by means of the following procedure. One starts from
arbitrary $N$ symbols, than the subsequent symbols are generated
with the probability that depends on the values of the preceding $N$
symbols only and is independent of the farther symbols,
\begin{equation}\label{defmar1}
P(s_i=s|T^-_{i,\infty})=P(s_i=s|T^-_{i,N}).
\end{equation}
Here $T^-_{i,L}$ is a set of $L$ sequential symbols
$(s_{i-L},s_{i-L+1},\dots,s_{i-1})$. The statistical properties of
such chains were studied in detail in
Refs.~\onlinecite{uya,muyag,allmemstepcor,highord,isotr,biased}.

Statistical equivalence of the Ising and Markov chains is widely
known (see, e.g., Refs.~\onlinecite{hen,bes}). However, these
References define Markov chains not via the ``one-sided''
conditional probability function Eq.~\eqref{defmar1}, depending on
the values of the preceding symbols, but rather via the
``two-sided'' one. The definition of the chain with one-sided
conditional probability function is more convenient for numerical
simulations of the chain as well as for obtaining analytic results.
Reference~\onlinecite{equiv} demonstrated the equivalence of the two
mentioned views on Markov chains.

The paper is organized as follows. In the second Section we show how
to find the Markov chain corresponding to the Ising chain under
consideration. Using this correspondence rule we calculate the RMS
value of the fluctuating part of magnetization for the chain
fragments of the length $L$ in the absence of magnetic field. We
also find magnetization of the chain in the presence of the magnetic
field. The third Section introduces the method for calculating the
statistical quantities of the fragments of Ising chain, in
particular, the internal and total energy, and entropy. All these
quantities are non-additive, i.e., they do not scale as the fragment
length $L$. We obtain analytical results for the two limiting cases
of low and high temperatures.

\section{Magnetic properties of Ising chain}

First, we find a one-sided conditional probability function of an
$N$-step Markov chain, Eq.~\eqref{defmar1}, that is statistically
equivalent to the Ising chain with Hamiltonian (\ref{hamilt}). Then
we use this property to find the magnetization of the Ising chain
for the case of weak interactions (compared to the temperature).
Finally, we study another limiting case of strong interactions.

\subsection{Equivalence of Markov and Ising chains}

The conditional probability of a spin to have a definite value, say
1, given that the values of all other spins in the chain are fixed,
is independent of the values of the spins separated by the distances
larger than $N$ (see, e.g., Refs.~\onlinecite{hen,bes}) and reads
\begin{equation}\label{zu}
P(s_i=1|T^-_{i,\infty},T^+_{i,\infty})=P(s_i=1|T^-_{i,N},T^+_{i,N}).
\end{equation}
Here $T^+_{i,L}$ is the $L$-tuple, $(s_{i+ 1},s_{i+ 2},\dots,s_{i+
L})$. The probability function in Eq.~(\ref{zu}) is given by the
Gibbs formula, and for the Hamiltonian \eqref{hamilt} it has the
following form:
\begin{equation}\label{uslver}
P(s_i=1|T^-_{i,N},T^+_{i,N})=\Big(1+\exp\big(-\dfrac{2H}{T}-\sum\limits
_{r=1}^{N}\dfrac{2\varepsilon
(r)}{T}(s_{i-r}+s_{i+r})\big)\Big)^{-1}.
\end{equation}

Expression~\eqref{uslver} for the two-sided conditional probability
function can be derived by the method proposed in Section III. It
can be shown that the Metropolis scheme~\cite{metro} also yields the
same expression~\eqref{uslver} for the conditional probability.

Reference~\onlinecite{equiv} analytically proved that the chain
defined by a two-sided conditional probability function (\ref{zu})
is equivalent to the $N$-step Markov chain defined by
Eq.~\eqref{defmar1}. The relation between the two-sided and
one-sided Markov conditional probability functions reads
\begin{equation}\label{dsv}
P(s_i=s \mid T^-_{i,N},T^+_{i,N})=\displaystyle\frac{
\prod\limits_{{r=0\atop s_i=1}}^N P(s_{i+r} \mid T^-_{{i+r},N})
}{\prod\limits_{{r=0\atop s_i=1}}^N P(s_{i+r} \mid
T^-_{{i+r},N})+\prod\limits_{{r=0\atop s_i=-1}}^N P(s_{i+r} \mid
T^-_{{i+r},N})}.
\end{equation}

Unfortunately, this relation is quite cumbersome and can hardly be
applied to the \emph{analytical} study of Ising chains in the
general case. Nevertheless, it allows one to calculate
\emph{numerically} the one-sided conditional probability function
for the $N$-step Markov chain. Thus, to attain equilibrium state of
the spin chain, we can generate the Markov chain according to the
algorithm described in the Introduction, instead of using the
Metropolis scheme.

\subsection{\label{weak} Magnetization of the
Ising chain with weak interaction}

In order to use Eqs.~\eqref{uslver} and \eqref{dsv} for the
analytical study of Ising chains we consider the case of \emph{weak
interaction} compared to the temperature,
\begin{equation}\label{small}
\sum\limits_{r=1}^N \varepsilon(r)\ll T.
\end{equation}
Expanding Eq.~\eqref{uslver} over the small parameter
$\sum\limits_{r=1}^N \varepsilon(r)/ T$ and using Eq.~\eqref{dsv} we
obtain the one-sided conditional probability function. In the first
approximation, it takes the additive form,
\begin{equation}\label{adm}
P(s_i=1 \mid T^-_{i,\infty}) =\dfrac{1+\bar{s}}{2}+\sum_{r=1}^{N}
F(r)(s_{i-r}-\bar{s}).
\end{equation}
Here the function $F(r)$ and $\bar{s}$ (the average value of $s_i$)
are determined as follows,
\begin{equation}\label{fe}
F(r)= \frac{\varepsilon (r)}{2T}\cosh^{-2}\frac{H}{T},\quad
\bar{s}=\Big(1+4\sum_{r=1}^NF(r)\Big)\tanh\frac{H}{T}.
\end{equation}
The Markov chain with additive conditional probability function,
Eq.~\eqref{adm}, is referred to as the \emph{additive} Markov chain.
The function $F(r)$ is referred to as the memory
function~\cite{muyag}. Equation (\ref{fe}) reflects the obvious fact
that an increase of the temperature leads to an increase of disorder
and to decrease of correlations in the chain. It should be noted
that the interaction range $N$ in Eqs.~(\ref{adm}) and (\ref{fe})
can be taken as $\infty$, if the series $\sum_r |F (r)|$ converges
and the value of $P$ in Eq.~(\ref{adm}) satisfies the inequality
$0<P(s_i=1 \mid T^-_{i,\infty})<1$.

Thus, the Ising chain in the equilibrium is equivalent to the
additive Markov chain. For the case of weak interaction between the
spins, the corresponding memory function $F(r)$ is proportional to
the energy $\varepsilon(r)$. The memory function of the additive
Markov chain is related to its pair correlation function $K(r)$,
\begin{equation}\label{11111}
K(r)=\overline{\phantom{Z}\!\!\!\!s_i s_{i+r}}-\bar{s}^2,
\end{equation}
by the recurrence equation~\cite{MUYa05},
\begin{equation}\label{k11}
K(r)=\sum_{r'=1}^{N} 2F(r')K(r-r'),\quad r=1,2,\dots.
\end{equation}
The overline $\overline{\phantom{Z}\!\!\!\!\!\dots}$ in
Eq.~(\ref{11111}) denotes statistical averaging. In the limiting
case of small $F(r)$, Eq.~\eqref{k11} yields approximately,
\begin{equation}\label{kf}
K(r)\approx 2F(r),\quad r=1,\dots,N.
\end{equation}

\subsubsection{Fluctuations of magnetization in the absence of
magnetic field}

We now consider an Ising chain with weak but long-range interaction
in the absence of the magnetic field. As mentioned above, the
averaged magnetization equals to zero in this case. However, the
root-mean-square (RMS) value of the magnetization of a chain segment
with length $L$,
\begin{equation}
M_{\rm
RMS}(L)=\Big(\overline{(s_{i+1}+s_{i+2}+\dots+s_{i+L})^2}\Big)^{1/2},
\end{equation}
is not zero and is determined by the correlation function of the
chain,
\begin{equation}\label{rms}
M_{\rm
RMS}(L)=\Big(L+2\sum\limits_{r=1}^{L-1}(L-r)K(r)\Big)^{1/2}.
\end{equation}

Using Eqs.~(\ref{fe}) and~(\ref{kf}), we derive the relation between
the RMS of magnetization and the interaction energy,
\begin{equation}\label{var}
M_{\rm RMS}(L)=\sqrt{L}+\dfrac{1}{T\sqrt{L}}
\sum\limits_{i=1}^{L-1}(L-i)\varepsilon(i)%
.
\end{equation}
Here we used that $\varepsilon(r)=0$ for $r>N$, thus the range $N$
of interaction does not appear explicitly.

The interaction between spins leads to the deviation of the RMS of
magnetization from the value $\sqrt{L}$ occurring in the
non-interacting chain. If this deviation is small, as we assume
here, the distribution function of magnetization is close to the
Gaussian form with the variance $M^2_{\rm RMS}(L)$.

\subsubsection{Magnetization in the external magnetic field}

If the magnetic field $H$ is applied, the averaged value $\mu$  of
magnetization (per spin) is no longer zero. In the assumption of
weak interactions, Eq.~\eqref{small}, $\mu$ is equal to the averaged
value of the spin $\bar{s}$, Eq.~\eqref{fe},
\begin{equation}\label{mu}
    \mu=\bar{s}=\Big(1+
    \frac{2}{T}\cosh^{-2}\frac{H}{T}
    \sum\limits_{r=1}^N \varepsilon(r)\Big)\tanh\frac{H}{T}.
\end{equation}
In the nearest-neighbor approximation $(N=1)$, this equation
coincides with the high-temperature asymptotics of the well-known
expression for the magnetization,
\begin{equation}\label{magnet}
    \mu={\rm sign}\, H\left(1+\frac{\exp(-4\varepsilon/T)}{\sinh^2(H/T)}
    \right)^{-1/2}.
\end{equation}
Here $\varepsilon=\varepsilon(1)$ is the single exchange integral
for the nearest neighbors.

\begin{figure}[t]
\begin{centering}
\scalebox{0.8}[0.8]{\includegraphics{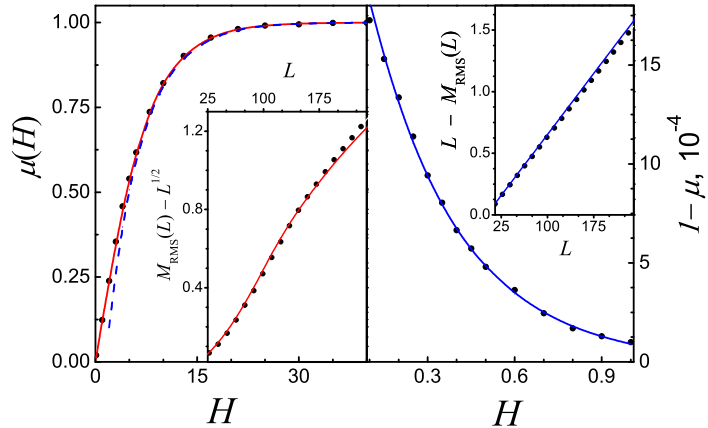}} \caption{(Color
online) The magnetization $\mu$ of the chain per spin versus the
magnetic field $H$. Left panel ($T$=10): solid and dashed curves
correspond to the high-temperature, Eq.~\eqref{mu}, and
low-temperature, Eq.~\eqref{mu1}, limits, respectively. Right panel
($T$=0.6): solid curve corresponds to low-temperature,
Eq.~\eqref{mu1}, limit. The RMS magnetization in the absence of
magnetic field versus length $L$ of a segment is presented in the
insets. The dots correspond to the results of the numerical
simulations. The function $\varepsilon(r)$ is step-like,
$\varepsilon(r)=1/100$ for $r =1,2,\dots,100$ and $\varepsilon(r)=0$
for $r >100$.} \label{murms}
\end{centering}
\end{figure}

The results of numerical simulations for the Ising chain with weak
interaction are given in the left panel of Fig.~\ref{murms}. Here we
present the dependence of the magnetization on the value of magnetic
field. The inset shows the dependence of the RMS of magnetization on
the segment length $L$ in the absence of the magnetic field. The
numerical points are very close to the corresponding analytical
curves (solid lines) determined by Eqs.~\eqref{mu} and~\eqref{var}.

\subsection{Magnetic properties of the Ising chain
in the low-temperature limit}

In the previous subsection, we have used the statistical equivalence
of the Markov and Ising chains for the calculation of the
magnetization in the case of high temperatures. Here we examine the
opposite limiting case of low temperatures. The corresponding
inequality imposed on the temperature $T$, the interaction-energy
function $\varepsilon(r)$, and magnetic field $H$ will be given
below (see Eq.~\eqref{ogr}). In this case, Eq.~\eqref{uslver}
defines a non-additive Markov chain (see Ref.~\onlinecite{equiv} for
details). Since finding the correlation function for such chains
brings about considerable technical difficulties, we examine the
required statistical properties of the Ising chain directly from
Eq.~\eqref{uslver}.

\subsubsection{Statistical properties of strongly correlated Ising
spin chain}

In the case of low temperatures, the conditional probability
determined by Eq.~\eqref{uslver} is close either to zero or one.
Indeed, due to strong correlations, the chain consists of large
domains predominantly occupied by the same symbols. It is evident
that, once strong enough magnetic field is applied, the chain will
mostly consist of the same symbols.

For definiteness, we consider the case when almost all symbols in a
chain are 1 (the case of $H>0$). In other words, the probability to
find the spin $-1$ at a given place of the chain is small. For
further convenience, it is suitable to introduce the parameter
$\delta$,
\begin{equation}\label{delta}
\delta=\exp (-4\epsilon_1/T)\ll 1, \quad
\epsilon_1=\sum\limits_{r=1}^N \varepsilon(r)+H/2.
\end{equation}
According to Eq.~\eqref{uslver}, this parameter is a ratio of the
probability to find the ($2N+1$)-tuple with only one spin $-1$ in
the center to the probability of occurring the ($2N+1$)-tuple with
all spins $1$. Due to the correlations, the probability to have the
($2N+1$)-tuple with one more spin $-1$ at distance $k$ from the
central spin $-1$ differs from $\delta^2$ and is equal to
$\delta^2\sigma_{k}$ with
\begin{equation}
\label{sigma}
\sigma_k=\exp (4\varepsilon(k)/T) \geqslant 1.
\end{equation}
Further, due to Eq.~\eqref{uslver}, three  spins $-1$ (at successive
distances $k$ and $l$) occur even less frequently, with the
probability $\delta^3\sigma_k\sigma_l\sigma_{l+k}$. The
corresponding probabilities for four and more spins $-1$ can be
found in a similar way.

We use the following assumptions for the parameters of the system:
\begin{equation}\label{ogr}
 \delta\sigma_1\sum_{l=1}^N\sigma^2_{l}
 \ll\big(1+\exp(-2H/T)\big)^{-2N}.
\end{equation}
This inequality states that the probabilities to have a
($2N+1$)-tuple with three or more spins $-1$ is negligible compared
to that with just two spins $-1$. Inequality (\ref{ogr}) holds only
for the case of monotonously decreasing functions $\varepsilon(r)$.
In general case, we should replace $\sigma_1\sum\sigma^2_{l}$ in
Eq.~\eqref{ogr} by the maximum value of
$\sigma_k\sum\sigma_l\sigma_{l+k}$.

Inequality~\eqref{ogr} can be considerably simplified under some
suppositions. For example, if function $\varepsilon(r)$ is rapidly
decreasing ($\varepsilon(1)\gg\varepsilon(2)\gg\dots$) and
$H>4\varepsilon(1)$ we obtain
\begin{equation}\label{ogr1}
   T\ll H-4\varepsilon(1).
\end{equation}
For chains with approximately step-like function $\varepsilon(r)$
($\varepsilon(r)\thickapprox\varepsilon(1)$ for $r\leq N$ and
$\varepsilon(r)=0$ for $r>N$), we have
\begin{equation}\label{ogr2}
   T\ll N\varepsilon(1)+H /\ln N.
\end{equation}
Thus, Eq.~\eqref{ogr} means that the temperature $T$ should be
sufficiently \emph{small} compared with $H$ or $\varepsilon(i)$.

It should be noted that inequalities~\eqref{ogr} and~\eqref{small}
can be satisfied simultaneously. This is the case when $\sum
\varepsilon(r) \ll T\ll H/N$. Then, the results of
subsection~\ref{weak} coincide with those of this subsection
(Eq.~\eqref{mu} transforms to Eq.~\eqref{mu1}).

\subsubsection{Magnetization of the Ising chain}

Using Eqs.~(\ref{uslver}), (\ref{delta}), and (\ref{sigma}), we
derive average magnetization $\mu$ per spin in the presence of the
magnetic field $H>0$,
\begin{equation}\label{mu1}
    \mu=\bar{s}=1-2\delta+2\delta^2-4\delta^2\sum_{k=1}^N(\sigma_k-1).
\end{equation}
For strong enough magnetic fields, $\exp(-H/T)\ll 1$, this
expression is also valid in the high-temperature limit
Eq.~(\ref{small}). Moreover, for chains with nearest-neighbor
interaction ($N=1$), it coincides with the general formula
\eqref{magnet} in the main and first approximations with respect to
the parameter $\exp(-(H-4\varepsilon)/T)\ll 1$.

If the magnetic field is not applied, the averaged magnetization is
equal to zero. However, the RMS of magnetization, determined by the
pair correlation function $K(r)$, is non-zero. In the case of low
temperatures, $K(r)$ is close to one. For $r \leqslant N$,
\begin{equation}\label{cor}
K(r)\approx 1-4\delta.
\end{equation}
Therefore, the RMS of magnetization of segments with length $L \ll
L_0$ is equal to $L$ in the main approximation with respect to
$\delta$, contrary to the square-root dependence $\sqrt{L}$ valid in
the high-temperature limit. Here $L_0 \gg N/\delta$ is the
characteristic length of the domains consisting of the same symbols,
1 or $-1$. Using Eqs.~(\ref{rms}) and~(\ref{cor}), we obtain,
\begin{equation}
M_{\rm RMS}(L)=L-2(L-1)\delta,
\end{equation}

The results of numerical study of strongly correlated chains are
shown in Fig.~\ref{murms}. As in the case of weak interaction, we
present calculations of the RMS of magnetization for the segment of
length $L$ in the absence of the magnetic field (inset in the right
panel) and the averaged magnetization as the function of the
magnetic field $H$ (solid curve in the left panel and dashed curve
in the right panel). All the results of numerical simulations are
close to the theoretical predictions. As mentioned above, two
asymptotics Eqs.~(\ref{mu}) and~(\ref{mu1}) coincide for the strong
enough magnetic fields at $\sum\varepsilon(r)\ll T$. This is clearly
seen in the left panel.

Thus, we have studied the magnetization and its RMS for the Ising
chains in the two limiting cases of high and low temperatures. These
quantities are determined by the statistical properties of the
chains, particularly, by the correlation function. Not only the
magnetization, but all the other statistical properties of the Ising
chain can be studied using the equivalence of the Ising and Markov
chains (see Ref.~\onlinecite{highord}). In the next section, we
propose a general way of calculating the statistical quantities of
the Ising chains with long-range spin interaction.

\section{Total and internal energy and entropy of
Ising chain segments}

Here we present an approach to calculate any statistical quantity
(energy, entropy, etc.) for a subsystem of arbitrary length $L$ (a
segment of $L$ sequential spins in the Ising chain). This
mesoscopical subsystem, denoted by $M$, interacts with the
thermostat and with the rest of the chain. We denote also two
``border'' subsystems of the length $N$ by $B$. All the other spins,
except for $M$ and $B$, are denoted by $R$:
\begin{widetext}
\[
\underbrace{\dots s_{i-N}}\limits_{R}\underbrace{s_{i-N+1} \dots
s_{i}}\limits_{B} \underbrace{s_{i+1} \dots
s_{i+L}}\limits_{M}\underbrace{s_{i+L+1}\dots
s_{i+L+N}}\limits_{B}\underbrace{s_{i+L+N+1}\dots}\limits_{R} \]
\end{widetext}

The total energy of the chain is not equal to the sum of the
internal energies of subsystems $M$ and ($B+R$). So, the long-range
interaction between $M$ and ($B+R$) makes it impossible to use
directly the standard methods of statistical physics for calculation
the statistical quantities for segment $M$.

In order to find the condition of equilibrium, we introduce the
statistical \emph{ensemble} of the chains. We choose the
\emph{subensemble} of chains with fixed border regions $B$ (but with
varying $M$ and $R$). For each of $2^{2N}$ such subensembles, the
subsystem $B$ plays the role of a separating wall. If the ensemble
is in the equilibrium state, the total energy of ($M+R$) (that
includes the sum of internal energies of $M$ and $R$, and energy of
their interaction with $B$) is constant for every subensemble. Thus,
we can use the equilibrium condition between $M$ and $R$:
\begin{equation}\label{equil_nonext}
\dfrac{\partial \ln W_M(E_M|B)}{\partial E_M}=\dfrac{\partial \ln
W_R(E_R|B)}{\partial E_R}=\frac{1}{T}\,,
\end{equation}
where $W_M(E_M|B)$ and $W_R(E_R|B)$ are the statistical weights of
subsystems $M$ and $R$ with total energies $E_M$ and $E_R$, with the
borders $B$ fixed. We refer to statistical weights $W_M(E_M|B)$ and
$W_R(E_R|B)$ as the \emph{conditional statistical weights}. It
should be emphasized that these weights are not mesoscopical
quantities because of their dependence on the microscopic states of
borders $B$. If the system is in the thermal contact with the
Gibbsian thermostat, the condition of equilibrium between the
thermostat and segment $M$ sets the temperature of $M$ equal to $T$.
This temperature is obviously the same for every subensemble with
fixed $B$. Thus, the averaged temperature of the subsystem $M$ is
also $T$.

Note that the external thermostat is not necessary for establishing
the temperature $T$ of the mesoscopical segment $M$. The role of a
thermostat can be played by the subsystem $R$. Indeed, the
temperature determined by Eq.~\eqref{equil_nonext} does not depend
on values of spins in $B$ (because the statistics of infinite
subsystem $R$ can not depend on the microstate of finite subsystem
$B$). This means that, for all the subensembles, the temperature of
$M$ is the same even in the absence of the external thermostat. In
this case, the temperature in the chain is determined by its initial
state (that can be nonequilibrium one).

It is very important that the distribution function of the segment
$M$ over different microstates \emph{within a subensemble with fixed
borders $B$} is Gibbsian (though the segment $M$ itself is not a
Gibbsian system). Therefore, within every subensemble, we can
introduce a conditional statistical quantity $Q_M(|B)$ for the
segment $M$. The actual quantity $Q_M$ is the conditional one,
averaged over the subensembles with different borders $B$:
\begin{equation}\label{Q}
Q_M=\big\langle Q_M(|B)\big\rangle_{B}\,\,.
\end{equation}

With this method for calculating the statistical quantities, we do
not need to find the distribution function for the segment $M$ over
different microstates in the whole ensemble. Note also that
Eq.~(\ref{Q}) can be considered in the thermodynamical limit $L\to
\infty$. However, we focus our attention on the mesoscopical
segments $M$ with finite length $L$.

The \emph{conditional entropy} can be introduced as the logarithm of
the conditional statistical weight: $S_M(E_M|B)=\ln W_M(E_M|B)$.
Equality $dS_M(E_M|B)=dE_M(T|B)/T$ is fulfilled for the
\emph{conditional} quantities $S_M(|B)$ and $E_M(|B)$. Meanwhile,
such a relation is not valid for the \emph{averaged} entropy
$S_M(E_M)=\big\langle S_M(E_M|B)\big\rangle_{B}$ and total energy
$E_M(T)=\big\langle E_M(T|B)\big\rangle_{B}$.

Note that the presented method for calculating the statistical
quantities is rather general and can be applied, e.g., to the
internal energy of a segment $M$ that can be measured
experimentally, or to the probability for some spin to take on the
definite value under the condition of fixed environment. Considering
one spin $s_i$ as a subsystem $M$ and the conditional probability
$P$ as a quantity $Q$ in the above-mentioned method, one arrives at
Eq.~\eqref{uslver}.

Now we apply our approach to calculation of the \emph{total} and
\emph{internal} energy and entropy of the segments of length $L$ for
the limiting cases of high and low temperatures.

\subsection{High-temperature limit}

The total energy of a segment $M$ with length $L$ for prescribed
configuration of $M$ and $B$ is expressed via products of pairs of
spins,
\begin{equation}\label{eee}
E(L|M,B)=-H\sum_{j=i+1}^{i+L}s_j
-\sum_{j=i+1}^{i+L}\sum_{\scriptstyle{k=j-N}\atop
\scriptstyle{k\not=j}}^{j+N} \varepsilon(|k-j|)s_k s_j +
\sum_{\scriptstyle{j,k=i+1}\atop \scriptstyle{j<k}}^{i+L}
\varepsilon(k-j)s_k s_j.
\end{equation}
The actual total energy is the conditional energy averaged over
the different configurations of $M$ and $B$,
$E(L,T)=\overline{E(L|M,B)}$. Along with the total energy
$E(L,T)$, the segment $M$ can be characterized by its internal
energy $E^{\rm in}(L,T)=\overline{E^{\rm in}(L|M,B)}$ specifying
the interaction between the spins in $M$ only,
\begin{equation}\label{eee1}
E^{\rm in}(L|M,B)=-H\sum_{j=i+1}^{i+L}s_j-
\sum_{\scriptstyle{j,k=i+1}\atop \scriptstyle{j<k}}^{i+L}
\varepsilon(k-j)s_k s_j.
\end{equation}
This quantity can be measured experimentally and thus is of interest
for study.

The second term in Eq.~(\ref{eee}) accounts the interaction of
each spin of the segment $M$ with $2N$ surrounding spins. The
internal interaction energy is accounted twice in this term. Thus,
the third term in Eq.~(\ref{eee}) subtracts the corresponding
extra summands.

In the high-temperature limit, the total and internal energy can be
calculated via the pair correlation function $K(r)$ with $r\leqslant
N$, without the need for the conditional energies. Using
Eqs.~(\ref{fe}) and~(\ref{kf}), we arrive at
\begin{gather}
E(L,T)=-\Big(\epsilon_{2}L-\sum\limits_{r=1}^{L-1}(L-r)
\varepsilon^2(r)\Big)/T, \label{energy}\\ E^{\rm in}(L,T)=
-\Big(\sum\limits_{r=1}^{L-1}(L-r)\varepsilon^2(r)+H^2 L\Big)/T,
\end{gather}
with
\begin{equation}\label{eps2}
\epsilon_{2}=2\sum\limits_{r=1}^N \varepsilon^2(r)+H^2,\quad H \ll
T.
\end{equation}
If the segment length $L$ is much greater than the memory length $N$
this expression yields additive energies, $E(L,T) \propto L$. For
the opposite limiting case, $L\ll N$, we get:
\begin{equation}
E(L,T) \approx -\epsilon_{2} L/T +\varepsilon^2(1)L^2/2T,\quad
E^{\rm in}(L,T) \approx -\varepsilon^2(1)L^2/2T-H^2L/T.
\end{equation}
Here we suppose that $\varepsilon(i)\thickapprox \varepsilon(1)$ for
$i\lesssim L$.

The non-additive energy is expressed in terms of the pair
correlation function only. This is not correct for other statistical
quantities, e.g., the entropy of the segment $M$. Formally, in order
to find the entropy, one should calculate all the conditional
entropies by \emph{integrating} the equation
$dS(L,T|B)=dE(L,T|B)/T$, and \emph{averaging} the result over all
realizations of the borders~$B$. However, at high temperatures, to a
first order in the small parameter $\epsilon_1/T$, we can change the
order of these operations and calculate the averaged entropy by
integrating the formula $dS(L,T)=dE(L,T)/T$ written for the averaged
energy. A constant of integration is determined from the condition
of complete randomization at high temperatures, $S(L,T\rightarrow
\infty) \rightarrow \ln \left(2^L\right)$. Thus, we obtain
\begin{equation}\label{entropy}
S(L,T)=L \ln 2+E(L,T)/2T
\end{equation}
with $E(L,T)$ given by Eq.~(\ref{energy}).

\begin{figure}[t]
\begin{centering}
\scalebox{0.8}[0.8]{\includegraphics{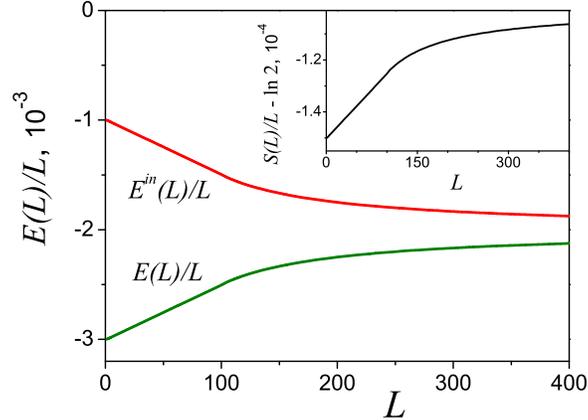}} \caption{(Color
online) The dependences of specific non-additive energy $E(L)/L$ and
$E^{\rm in}(L)/L$ on the length $L$ of a segment for the step-like
function $\varepsilon(r)=1/100, \quad r\leq 100$. The specific
non-additive entropy $S(L)/L$ versus length $L$ is presented in the
inset. Other parameters are: $T=10$, $H=0.1$. \label{esht}}
\end{centering}
\end{figure}

Expressions~(\ref{energy})--\eqref{entropy} describe the
non-additive dependence of energy and entropy of a segment $M$ on
its length $L$. Note that the energy is non-additive in the main
approximation in the parameter $\epsilon_1/T$, while the
non-additivity of the entropy appears in the first-order correction
only. The dependences of the non-additive energy and entropy on $L$
are presented in Fig.~\ref{esht} for the step-like function
$\varepsilon(r)$.

Knowing the energy and entropy we can find some other statistical
quantities. For example, at high temperatures the heat capacity can
be determined in a similar way as for the entropy. One can use the
classical formula $C(L,T)=TdS(L,T)/dT$ with averaged entropy given
in Eq.~\eqref{entropy} and obtain the equation
\begin{equation}\label{CM}
C(L,T) = -E(L,T)/T =
\Big(\epsilon_{2}L-\sum\limits_{r=1}^{L-1}(L-r)
\varepsilon^2(r)\Big)/T^2.
\end{equation}
This formula is valid for the case of high temperatures in the main
approximation only. In the general case, $C(L,T)$ is not determined
by equation $C(L,T)=TdS(L,T)/dT$. This relation holds for the
conditional quantities only.

For the case of high temperatures, we calculated the averaged
non-additive energy and entropy without using the conditional ones.
In the opposite limiting case of low temperatures, it is necessary
to calculate the conditional quantities as well.

\subsection{Low-temperature limit}

In this subsection, we assume that inequality Eq.~\eqref{ogr} is
fulfilled. Let us find the averaged values of the total and internal
energy and entropy of a segment $M$ consisting of $L$ spins. These
are close to their minimal values taken at $T=0$ when all spins in
segment $M$ (as well as in borders $B$) are oriented along the
magnetic field $H$. For positive magnetic fields, $H>0$, we have
\begin{equation}\label{enrgymin}
E_{\rm min}(L)=-(\epsilon_1+H/2)L- \sum\limits_{r=1}^N
r\varepsilon(r), \quad E^{\rm in}_{\rm
min}(L)=-(\epsilon_1+H/2)L+\sum\limits_{r=1}^N r\varepsilon(r).
\end{equation}
For simplicity, we consider here and below the segments with
$L\geqslant N$.

According to the Nernst theorem, the entropy in the zeroth
approximation is $S(L,T=0)=0$. Now we seek the first non-additive
corrections to $E$, $E^{\rm in}$, and $S$. Following the proposed
approach, we have to find, as a first step, the conditional total
energy and entropy for all the configurations of borders $B$:

i) The most probable configuration of borders $B$ is $B_0$ when they
consist of spins $+1$ only. The corresponding probability $P(B_0)$
is approximately equal to $1-2N\delta$. The conditional total energy
and entropy for this case can be found by virtue of
Eqs.~\eqref{delta},~\eqref{sigma}, and~\eqref{eee}:
\begin{gather}\label{ecse1}
E(L,T|B_0)=E_{\rm min}(L)+4\delta\epsilon_1L(1-L\delta)
+\delta^2\sum\limits_{k=1}^{L}\big(8\epsilon_1-4\varepsilon
(k)\big)(L-k) \sigma_{k} ,\\\notag S(L,T|B_0)=\int \limits_0^T
\dfrac{1}{T} dE(L,T|B_0)\\=\delta L\frac{4\epsilon_1+T}{T}-
\delta^2L^2 \frac{8\epsilon_1+T}{2T}
+\delta^2\sum\limits_{k=1}^{L}\frac{8\epsilon_1-4\varepsilon
(k)+T}{T}(L-k)\sigma_{k}.
\end{gather}

ii) Much less probable configuration of the borders is $B_{1,k}$
with only one symbol $-1$ at distance $k$ from the edge of segment
$M$. The corresponding probability does not depend on $k$ and is
approximately equal to $\delta$, $P(B_{1,k})\approx \delta$. For
this configuration, the conditional total energy and entropy are:
\begin{gather}\label{ecse2}
E(L,T|B_{1,k})=E_{\rm min}(L)+2\sum\limits_{j=1}^{L}\varepsilon
(k+j-1) +\delta\sum\limits_{j=1}^{L}\big(4\epsilon_1-4\varepsilon
(k+j-1)\big)\sigma_{k+j-1} ,\\\notag S(L,T|B_{1,k})=\int
\limits_0^T \dfrac{1}{T} dE(L,T|B_{1,k})= \delta
\sum\limits_{j=1}^{L} \dfrac{4\epsilon_1-4\varepsilon(k+j-1)+T}{T}
\sigma_{k+j-1}.
\end{gather}

iii) Probabilities of all other configurations of the borders are
negligible with respect to $P(B_{1,k})$. There is no need to find the
corresponding conditional energies and entropies if we are
interested in the first non-additive corrections to these
quantities.

Averaging the conditional quantities over border configurations
$B_{0}$ and $B_{1,k}$ and keeping the non-additive temperature
dependent terms in the main approximations in $\delta$, we derive
the actual quantities $E(L,T)$ and $S(L,T)$:
\begin{gather}
\label{averE} E(L,T)=E_{\rm min}(L)+ 4\delta\Big(\epsilon_1
L+\sum\limits_{r=1}^{N }r\varepsilon (r) \Big), \\
 S(L,T)=
L\delta\Big[\dfrac{4\epsilon_1}{T}(1-2\delta)+1+
\delta\frac{4H-T}{2T}+
 \delta\sum\limits_{k=1}^N
 \dfrac{8\epsilon_1-4\varepsilon(k)+T}{T}(\sigma_{k}-1)\Big]\\
 +\delta^2\Big[\sum\limits_{k=1}^N
 k(\sigma_{k}-1)-\frac{4}{T}\sum\limits_{k=1}^N
 k\varepsilon(k)\sigma_{k}\Big].
 \label{averS}
\end{gather}

Similar calculations for the internal energy give the following
expression:
\begin{equation}\label{enrgyi}
E^{\rm in}_{M}(L,T)=E^{\rm in}_{\rm
min}(L)+4\delta\Big(\epsilon_1L-\sum\limits_{r=1}^N
r\varepsilon(r)\Big).
\end{equation}

\begin{figure}[t]
\begin{centering}
\scalebox{0.8}[0.8]{\includegraphics{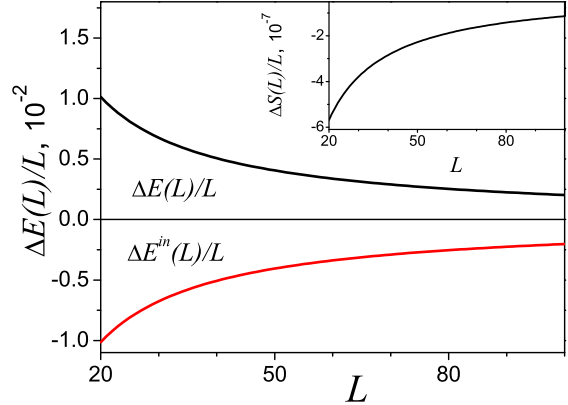}} \caption{(Color
online) The specific non-additive energies $E$, $E^{\rm in}$ and
entropy $S$ versus the length $L$ of a segment for step-like
function $\varepsilon(r)=1/20, \quad r\leq 20$. Symbol $\Delta$
denotes the non-additive terms in Eqs.~\eqref{averE},~\eqref{averS}
and~\eqref{enrgyi}. Other parameters are: $T=3$, $H=6$.
\label{eslt}}
\end{centering}
\end{figure}

One can see that the non-additive total energy and entropy,
determined by Eqs.~\eqref{averE} and~\eqref{averS}, do not satisfy
the relation $dE(L,T)=TdS(L,T)$, unlike the conditional total energy
and entropy. The dependences of the non-additive energy and entropy
on $L$ are presented in Fig.~\ref{eslt} for a step-like function
$\varepsilon(r)$.

\begin{figure}[t]
\begin{centering}
\scalebox{0.8}[0.8]{\includegraphics{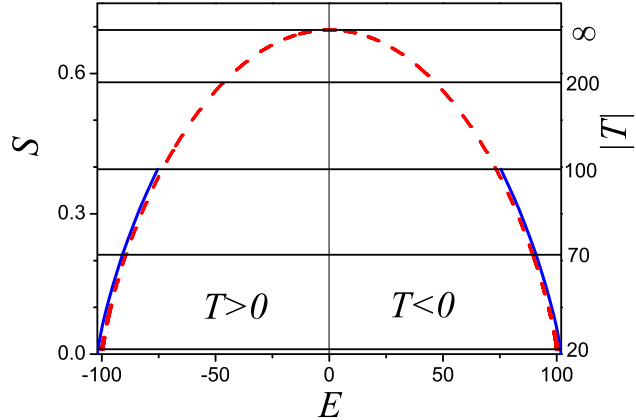}} \caption{(Color online)
The dependence of the entropy $S$ on the total energy $E$ for the
step-like function $\varepsilon(r)=1/20, \quad r\leq 20$. Magnetic
field $H=100$ is so strong that the high-temperature (red dashed
curve), Eqs.~(\ref{energy}),~(\ref{entropy}), and low-temperature
(blue solid curve), Eqs.~(\ref{averE}),~(\ref{averS}), asymtotes
coincide. The corresponding values of the temperature are shown on
the right scale. \label{diag}}
\end{centering}
\end{figure}

As mentioned above, in the absence of an external thermostat, the
temperature $T$ of a chain is controlled by its initial state. In
this case, the temperature can be expressed via the total energy of
the chain. In particular, the positive initial energy of the chain
provides \emph{negative} temperature (see Fig.~\ref{diag}). If the
initial energy is close to its maximum, the temperature is negative
and close to zero and the system is almost ``frozen''. The lesser
the positive energy the lower the negative temperature. If the
initial energy tends to zero, the chain becomes more chaotic and its
temperature tends to the minus infinity.

The situation is principally different for the chains in contact
with an external thermostat. The states of chain with negative
temperatures become unstable and, therefore, do not exist. If the
initial energy of the chain is negative and close to zero, the
temperature is positive and high. At the same time, it tends to zero
for the energy close to its minimum. Both with and without the
external thermostat, all the statistical properties of the chain
with positive temperature are the same.

\section{Conclusion}

Thus, we have studied the statistical properties of mesoscopic
segments of Ising spin chains with finite but arbitrary long
interaction ranges. The equivalence of the Ising and $N$-step Markov
chains was used for calculating the averaged statistical quantities.
In particular, the averaged magnetization in the presence of the
external magnetic field and its RMS were calculated in the two
limiting cases of strong and weak interactions. Correlations in the
chain result in the non-additive behavior of these quantities as the
segment length $L$ increases. We show that the statistical
quantities of the chain can be obtained by averaging the
corresponding conditional quantities. The explicit expressions for
the non-additive energy, internal energy, and entropy are derived in
the limiting cases of high and low temperatures comparing to the
energy of spin interaction. At high temperatures, the equilibrium
Ising chain of spin turns out to be equivalent to the additive
multi-step Markov chain.

\end{document}